%% file: expectation-entropy.tex
\documentclass[conference ]{IEEEtran}
\usepackage{cite, url}
\usepackage{graphicx, xcolor, svg}
\usepackage[linesnumbered, ruled]{algorithm2e} 
\usepackage{amsmath,amssymb, bm, mathrsfs}
\usepackage{enumerate} 
\usepackage[T1]{fontenc}
\usepackage{makecell}
\usepackage{siunitx}
\usepackage{tcolorbox}
\usepackage{tabularx,booktabs}
\usepackage[left=0.625in,right=0.625in,top=0.75in,bottom=1.0in]{geometry}

\SetCommentSty{mycommfont}
\IEEEoverridecommandlockouts

\graphicspath{ {images/} }

\begin{document}


\title{Expectation Entropy as a Password Strength Metric}


\author{
\IEEEauthorblockN{Khan Reaz, Gerhard Wunder}
\IEEEauthorblockA{Cybersecurity and AI Research Group\\
    Freie Universit\"at Berlin, Germany \\
    Email: khanreaz@ieee.org, g.wunder@fu-berlin.de
    }
    \thanks{This work is carried out within \textit{``PHY2APP: Erweiterung von Physical Layer Security für Ende-zu-Ende Absicherung des IoT"} project, which is funded by the German Federal Ministry of Education and Research (BMBF) under grant number 16KIS1473~\cite{phy2app}}
}
        
\maketitle


\begin{abstract}
\input{sections/0_abstract}
\end{abstract}
\begin {IEEEkeywords}
Password, Entropy, Randomness, NIST Entropy Estimation Tool
\end{IEEEkeywords}

\input{sections/1_section}

\input{sections/2_section.tex}
\input{sections/4_section.tex}

\input{sections/5_section}

\bibliographystyle{IEEEtran}
\bibliography{expectation-entropy.bib}

\end{document}

%% file: sections/0_abstract.tex
The classical combinatorics-based password strength formula provides a result in tens of bits, whereas the NIST Entropy  Estimation Suite give a result between 0 and 1 for Min-entropy. In this work, we present a newly developed metric -- \textit{Expectation entropy} that can be applied to estimate the strength of any random or random-like password.  \textit{Expectation entropy} provides the strength of a password on the same scale as an entropy estimation tool. Having an  \textit{Expectation entropy} of a certain value, for example, 0.4 means that an attacker has to exhaustively search at least 40\% of the total number of guesses to find the password.

%% file: sections/1_section.tex
\section{Motivation}
\label{sec:expectation-entropy-motivation}

The \textit{recipe} for constructing a strong password has two ingredients: randomness, and  length, i.e., the characters of the password must have high randomness (ideally truly random) and  the number of characters should be large. A good combination of these properties can make  brute-force attacks infeasible for that password.

State-of-the-art password strength estimation methods can be grouped into three main categories~\cite{galbally2016}: (1) Attack-based methods estimate the time it takes to break it with brute force. (2) Heuristic-based methods are based on Shannon's notion of entropy bits. (3) Probabilistic-based methods consider human intrinsic nature, context, and imposed password composition rules. NIST's Entropy Estimation  Suite~\cite{entropyassesment} is the industry  standard solution to estimate min-entropy using 10 tests as described in~\cite{nistrandomentropy2018}, \cite{burr2013nist}. 

In our previous work~\cite{reazComPassIMIS2021}, the device provisioning problem of Wi-Fi personal mode has been addressed and provided with \textit{ComPass} protocol to supplement WPA2/WPA3. \textit{ComPass} foregoes the pre-printed or user-generated password with an automatically generated strong symmetric password for the participating devices. The generated password is between 16 and 32 characters long and is generated by extracting signal parameters from typical Wi-Fi OFDM signals using Physical Layer Security methods. As we wanted to analyse  the randomness of  the quantise bit-string, and the strength of the generated password, we noticed that the classical combinatorics-based password strength formula: $\log_{2}(\text{character space}^{\text{length of the password}})$ provides the result in tens of bits, whereas entropy estimation formulae (and NIST entropy estimation tool) give  a result between 0 and 1.

In this work, we present a newly developed metric -- \textit{Expectation entropy} that can be applied to estimate the strength of any random or random-like password. It captures the composition of the characters and estimate the  strength from a single password. \textit{Expectation entropy} provides the strength of the password on the same scale as entropy estimation formulae and NIST Entropy Estimation Suite.

%% file: sections/2_section.tex
\section{Various Definitions of Entropy}

Entropy is  a measure of the disorder, randomness or variability in a closed system. The larger the amount of entropy, the greater the uncertainty in predicting the value of an observation. It is usually denoted by $ H $.  If $p_i$ is the probability of an element of a random discreet variable, then the min-entropy, $H_{\infty} $  is the largest value $ m $ having the property that each observation of the variable guarantees at least $ m $ bits of information: $H_{\infty} = - \log_{2} (max(p_{i})  )$.

Shannon entropy, $H_1$ (or just $H$) was introduced by Claude Shannon in \cite{shannon1951prediction} as : $ H_{1} = - \sum_{i = 1}^{N} p_{i} \log_{2}pi $, where $N = 2$ for binary digits, and  $N = 26$ for  English letters. Shannon entropy gives an average estimate and ignores the length of the password. Ralph Hartley proposed a quantitative measure of ``information'' two decades prior to Shannon \cite{hartley1928transmission}. Hartley entropy measures only the size of the distribution and ignores the probabilities: $ H_{0} = \log_{2} N$.



However, in a password cracking case, an attacker must guess each value one at a time which makes all previously  mentioned metrics unrelated to guessing difficulty~\cite{bonneau2012thesis}. Massey~\cite{massey1994guessing}, Cachin~\cite{cachin1997entropy}, and Pliam~\cite{pliam2000incomparability} individually developed the \textit{Guessing entropy} concept which  states the expected number of guesses required  if the attacker optimally tries: $ G = \sum_{i=1}^{N} p_{i} \cdot i   $.

%% file: sections/4_section.tex
\section{Expectation entropy}

Let us define four disjoint element sets such that $\mathcal{L} $ be the set of lower-case letters, $\mathcal{U}$ be the set of upper-case letters, $\mathcal{D}$ be the set of digits, and $\mathcal{S}$ be the set of symbols. Which means, $|\mathcal{L}| = 26$, $|\mathcal{U}| = 26 $, $|\mathcal{D}| = 10 $, $|\mathcal{S}| = 32$, and their disjoint union resulting in the total character space $\mathcal{K}$ of cardinality $|\mathcal{K}| = |\mathcal{L}|+ |\mathcal{U}| + |\mathcal{D}| + |\mathcal{S}|= 94$ for English language. Naturally, one can choose the character space for  a different  language.

A password $P$, where each of its character $c$ is chosen uniformly at random, is called valid if it contains characters from at least two disjoint element sets, and the password length $|P|$ satisfies $ min(|\mathcal{L}| , |\mathcal{U}| , |\mathcal{D}|, |\mathcal{S}| ) \leq |P|$. Then the probability of a character from each  element sets would be: $p_{\mathcal{L}} =  \mathbb{P}( c \in \mathcal{L}) = \frac{26}{94} $, $p_{\mathcal{U}} =  \mathbb{P}( c \in \mathcal{U}) = \frac{26}{94} $, $p_{\mathcal{D}} =  \mathbb{P}( c \in \mathcal{D}) = \frac{10}{94} $, and $p_{\mathcal{S}} =  \mathbb{P}( c \in \mathcal{S}) = \frac{32}{94} $.

If $ l, u, d, s $ are the  number of characters   chosen at random from  $\mathcal{L} $, $\mathcal{U} $, $\mathcal{D} $, $\mathcal{S} $ sets respectively such that $l, u, d, s \leq |\mathcal{K}|$. We find the  expectation of a character $c$ appearing in a password as:
\begin{equation}
 E(c(P)) = p_{\mathcal{L}} \cdot l + p_{\mathcal{U}} \cdot u + p_{\mathcal{D}} \cdot d + p_{\mathcal{S}} \cdot s  
\end{equation}
 The  maximum entropy (which is traditionally defined as Hartley entropy) of the total character space, $\mathcal{K}$ is $H_0(\mathcal{K}) = \log_{2} |\mathcal{K}|$.  Then we express the \textit{Expectation entropy} $H_E$ of the password $P$ as:

\begin{equation}
\label{eq:1}
    H_E(P) = \frac {\log_{2} E(c(P))}{H_{0}(\mathcal{K})} 
\end{equation}

It is not difficult to show that the upper bound is achieved for the random variable $E(c(P))$ if and only if $l, u, d, s$ are equal to $4 \cdot |\mathcal{K}|$. When the password length is larger than the restriction, $H_{E}$ gives a value more than 1, and for the smaller length it gives a negative value.

%% file: sections/5_section.tex
\section{Results}

Empirical evaluation has been done  using two different types of datasets. In the first case, 100,000 random passwords are generated  using a computer with a dedicated hardware random number generator. These passwords are labelled with \texttt{RandomMin}, \texttt{Random10ch}, \texttt{Random32ch}, \texttt{Random128ch}, and \texttt{RandomMax} based on the length of the password. In the second case,  three publicly leaked password databases (LinkedIn, 10Million, and WPA2) were used for evaluation.  

\begin{figure}[ht!]
\centering
    \includegraphics[width=\linewidth]{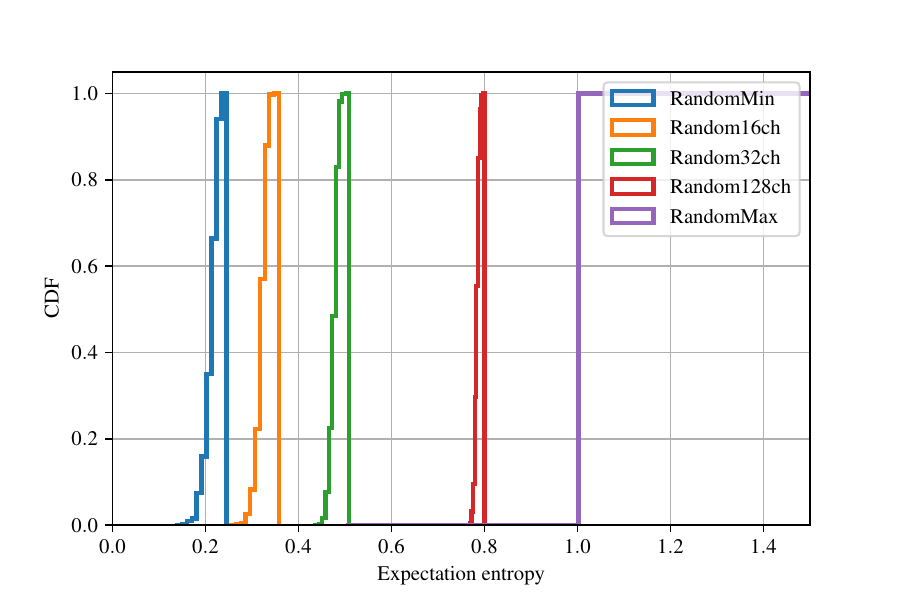}
    \caption{Comparison of Expectation entropy for randomly generated passwords with different length.}
    \label{fig:expectation_entropy_random}
\end{figure}

\begin{figure}[ht!]
\centering
    \includegraphics[width=\linewidth]{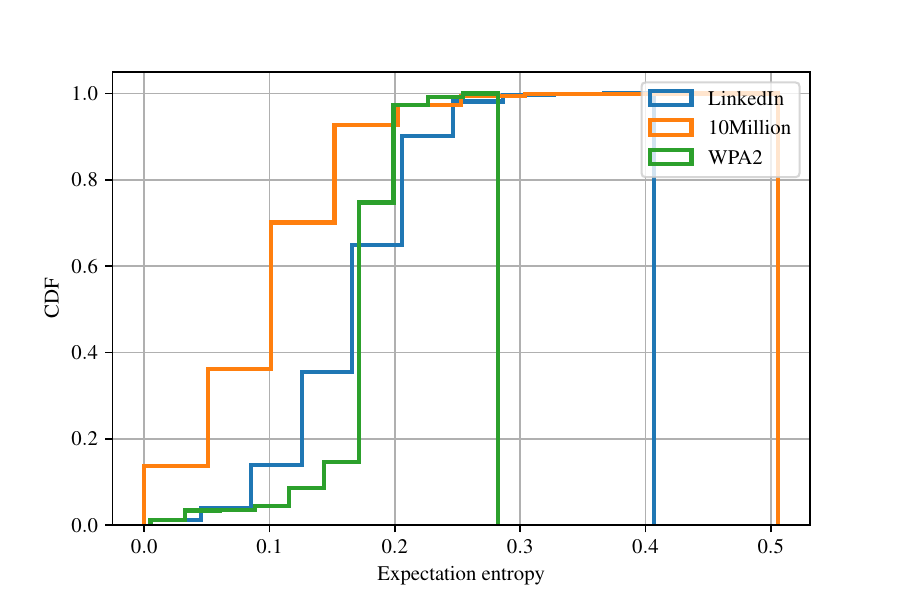}
    \caption{Comparison of Expectation entropy for leaked passwords.}
    \label{fig:expectation_entropy_leaked}
\end{figure}

Fig.~\ref{fig:expectation_entropy_random}, and Fig.~\ref{fig:expectation_entropy_leaked} summarise the result in terms of Cumulative Distributive Function (CDF) and \textit{Expectation entropy}. It can be observed from Fig.~\ref{fig:expectation_entropy_random} that the value of $H_{E}$ increases or decreases according to the length of the password and satisfies the bounds. Fig.~\ref{fig:expectation_entropy_leaked}  shows that the publicly leaked databases mostly contain  passwords with a short length and the characters of the password are not chosen from all element sets, hence low \textit{Expectation entropy}. Having an  \textit{Expectation entropy} of a certain value, for example, 0.4 means that an attacker has to exhaustively search at least 40\% of the total number of guesses to find the password using brute-force.